# High-accuracy absolute distance measurement by two-wavelength double heterodyne interferometry with variable synthetic wavelengths

Yoshiyuki KURAMOTO[*] and Hiroshi OKUDA

[1]*Corporate R &D Headquarters, Canon Inc., 23-10 Kiyohara-Kogyodanchi, Utsunomiya, Tochigi 321-3298, Japan*
[*]*Corresponding author: kuramoto.yoshiyuki@canon.co.jp*

We present an absolute distance measurement interferometer based on a two wavelength interferometer and a variable synthetic wavelength technique. The wavelength scanning range was 12 GHz, realized with a phase accuracy of 1.0 m$\lambda$ by heterodyne detection at each measurement wavelength. This small wavelength scanning range enabled the use of distributed feedback laser diodes as an interferometer light source and a fast 20 ms wavelength scanning time by injection current control. We demonstrated a measurement range of up to 1.5 m and an accuracy better than 1.2 nm in comparison with a displacement measurement interferometer, corresponding to a relative accuracy of $10^{-9}$. In addition, we also proposed expanding the range of maximum measurement and compensation of refractive index of air for linear colliders.
*2013 LCWS, 11-15 November, The University of Tokyo*

## 1. Introduction

Conventional displacement measurement interferometers (DMI) typically used for stage positioning are highly precise. A DMI measures the displacement from a previously set position. This limitation makes it difficult to use a DMI for distance measurements between parts of large, immobile devices. Nonetheless, a high-precision absolute distance meter is required in industrial and scientific applications.

Methods for absolute distance measurements using multi wavelength interferometry or wavelength scanning interferometry have been proposed [1–8]. Applying optical frequency comb (OFC) technology to length metrology is an active research area [8–11]. Minoshima et al achieved 490 nm accuracy at 1 m using self-beat signals of fiber-based frequency combs as a high stability and high frequency time-of-flight signal [9]. Salvadé et al achieved 8 nm accuracy at 0.8 m using an OFC as a wavelength reference for the synthetic wavelength of a multi-wavelength interferometer [11]. OFC technology has achieved a relative accuracy of up to $10^{-8}$, but has had a limited application in distance measurements due to system complexity. A simpler, high precision system was proposed by Pollinger et al used dual potassium-referenced external cavity diode lasers, leading to an accuracy of 12 $\mu$m at 20 m, a relative accuracy of $6 \times 10^{-7}$ [12].

Here we propose new absolute distance measurement interferometer (ADMi) with a simple design and high accuracy. The interferometer consists of two $C_2H_2$ referenced distributed feedback laser diodes for optical communications. We demonstrated a relative accuracy down to $10^{-9}$ at a 1.5 m measuring distance.

## 2. Measurement principle

The ADMi is based on a two wavelength interferometer. Absolute distance is measured with the accuracy of a conventional DMI by determining the interferometric order of the measurement wavelength with a variable synthetic wavelength [11].

When the phase of distance measurement interference changes from $\phi_2$ to $\phi_3$ with a change in measurement wavelength from $\lambda_{11}$ to $\lambda_{12}$, the geometrical path difference $L$ is:

$$L = \frac{1}{2}\frac{\Lambda_{23}}{n_{g23}}(\phi_3 - \phi_2), \qquad (1)$$

with

$$\Lambda_{ij} = \frac{\lambda_i \lambda_j}{\lambda_i - \lambda_j}$$
$$n_{g\,ij} = n(\lambda_j) - \lambda_j \frac{n(\lambda_i) - n(\lambda_j)}{\lambda_i - \lambda_j}, \qquad (2)$$

where $\Lambda_{ij}$ is the synthetic wavelength with optical wavelength $\lambda_i$ and $\lambda_j$, $n_{g23}$ is the group index of refraction at $\lambda_i$ and $\lambda_j$. Absolute distance is thus measured by a phase variation from $\phi_2$ to $\phi_3$, with an infinite ambiguity range.

Wavelength scanning interferometry has two problems. First, the wavelength scanning range is generally low. A wide wavelength scanning range is required for high accuracy distance measurements since measurement accuracy is proportional to wavelength. Second, measurement error for moving objects is generally large. Scanning wavelength induce measurement errors lead to distance variations proportional to the magnification ratio of the center wavelength and wavelength scanning range. A fixed wavelength interferometer is effective at solving these problems [11]. By implementing an interferometer with a fixed wavelength $\lambda_1$, the geometrical path difference $L$ is a function of the synthetic wavelength $\Lambda_{12}$ and wavelength $\lambda_1$ as:

$$L = \frac{1}{2}\frac{\Lambda_{12}}{n_{g12}}(M + \phi_2 - \phi_1) = \frac{1}{2}\frac{\lambda_1}{n(\lambda_1)}(N + \phi_1), \qquad (3)$$

where $M$ is the interferometric order of the synthetic wavelength $\Lambda_{12}$, and $N$ is the interferometric order of wavelength $\lambda_1$. $M$ and $N$ are determined as:

$$M = \text{round}\left(\frac{n_{g12}}{n_{g23}}\frac{\Lambda_{23}}{\Lambda_{12}}(\phi_3 - \phi_2) - (\phi_2 - \phi_1)\right)$$

$$N = \text{round}\left(\frac{n_1}{n_{g12}}\frac{\Lambda_{12}}{\lambda_1}(M + \phi_2 - \phi_1) - \phi_1\right), \quad (4)$$

where $\text{round}()$ rounds the argument to the nearest integer. In order to determine the interferometric orders correctly, errors in the arguments of the round functions should be less than 1/2. These conditions are expressed as:

$$2\left(\frac{\Lambda_{23}}{\Lambda_{12}}+1\right)^2 \cdot d\phi^2 + \left(\frac{L}{\Lambda_{12}}\right)^2 \cdot \left(\left(\frac{d\Lambda_{23}}{\Lambda_{23}}\right)^2 + \left(\frac{d\Lambda_{12}}{\Lambda_{12}}\right)^2\right) < \left(\frac{1}{2}\right)^2$$

$$2\left(\frac{\Lambda_{12}}{\lambda_1}+1\right)^2 \cdot d\phi^2 + \left(\frac{L}{\lambda_1}\right)^2 \cdot \left(\left(\frac{d\Lambda_{12}}{\Lambda_{12}}\right)^2 + \left(\frac{d\lambda_1}{\lambda_1}\right)^2\right) < \left(\frac{1}{2}\right)^2$$

(5)

where $d\Lambda_{ij}$ and $d\lambda_i$ are wavelength accuracies of the synthetic wavelength $\Lambda_{ij}$ and wavelength $\lambda_i$, respectively, and $d\phi$ is the phase measurement accuracy. We assumed that phase accuracies for each wavelength have the same magnitude and are uncorrelated. Both inequalities in (5) are composed of a phase accuracy term and a wavelength accuracy term. There is a tradeoff between the wavelength scanning range, which is inversely proportion to $\Lambda_{23}$, and phase accuracy since the phase accuracy induced interferometric error is magnified by $\Lambda_{23}/\Lambda_{12}$ or $\Lambda_{12}/\lambda_1$.

Based on the above principle, we minimized the wavelength scanning range to 12 GHz with a high accuracy phase measurement of 1 m$\lambda$ using independent heterodyne detection for each wavelength. The short wavelength scanning range enables fast wavelength scanning by injection current control using simple and low-cost laser chips such as distributed feedback laser diodes for optical communications as light sources for the interferometer.

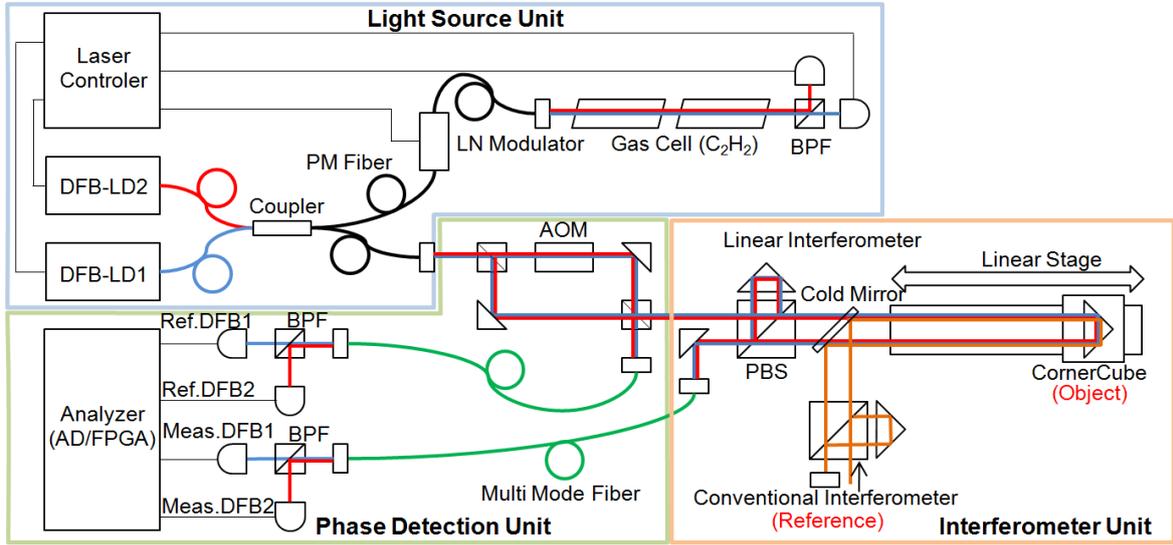

Fig. 1. (Color Online) Experimental setup for absolute distance measurement interferometer. Light source unit: Two DFB-LDs with injection current controlled wavelengths and $C_2H_2$ standard cells. Phase detection unit: Interference signals for each wavelength was divided by a band pass filter (BPF). Interferometric phases were detected independently by the heterodyne method. Interferometer unit: A linear interferometer was configured with a 1 m stroke linear stage and 200 μm stroke PZT stage for compensating cyclic error. For compensating optical pass length fluctuations, a conventional laser interferometer was coupled to a cold mirror.

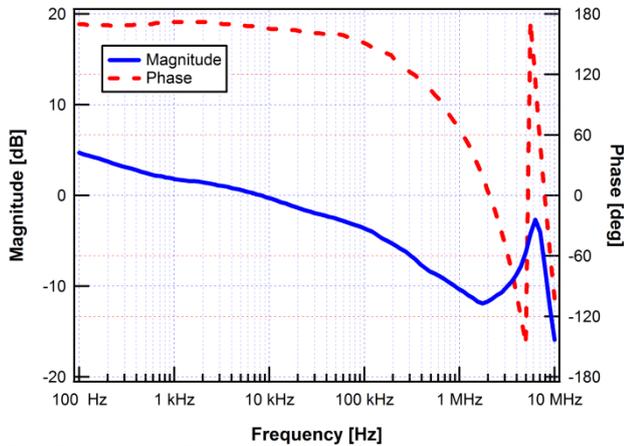

Fig. 2. (Color Online) Measured frequency response of DFB-LD.

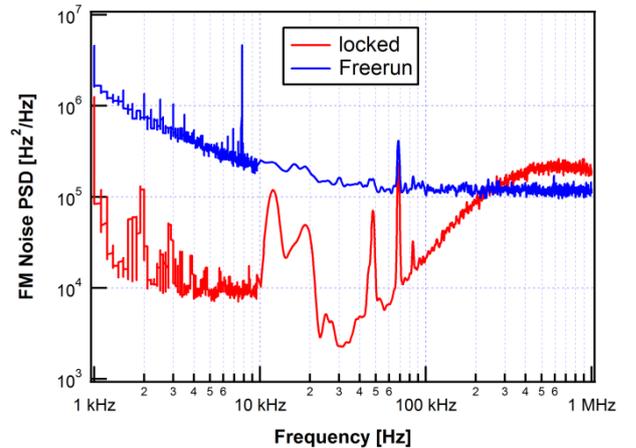

Fig. 3. (Color Online) FM noise spectra of DFB-LD measured for freerun and locked conditions to the $C_2H_2$ absorption line with a feedback control bandwidth of 200 kHz.

## 3. Experimental Setup

The experimental setup of the ADMi is depicted in Fig. 1. The ADMi consists of a light source unit, phase detection unit, and interferometer unit.

*Light Source Unit*

Two distributed feedback laser diodes (DFB-LDs) with a narrow linewidth of about 2 MHz (FRL15DCWA, FITEL) were used as light sources. Optical frequency sensitivity for the injection current of the DFB-LD was 457 MHz/mA. Output power sensitivity for the injection current was 0.23 mW/mA. We limited the range of injection current control for a wavelength scan by 60% of the specified maximum output power of 10mW. The maximum frequency scanning range of the DFB-LD was thus about 12 GHz.

We adopted absorption lines of two acetylene isotopes as a wavelength reference with a minimum frequency separation of 12 GHz. The DFB-LD1 was stabilized at 1539.97 nm ($\lambda_1$) referenced to $^{13}C_2H_2$ P2. DFB-LD2 was scanned between 1529.27 nm ($\lambda_2$) and 1529.18 nm ($\lambda_3$) and referenced to $^{13}C_2H_2$ R7 and $^{12}C_2H_2$ P7. These three wavelengths created two synthetic wavelengths of 218.2 μm ($\Lambda_{12}$) and 24.8 mm ($\Lambda_{23}$).

The two DFB-LDs were coupled using a 2x2 single mode (SM) polarization maintained (PM) 3 dB coupler. Half of the light was used to control wavelength, and half was used as a light source for the interferometer. The laser output for controlling wavelength was passed through a fiber coupled waveguided LN phase modulator (PM-150-005, JDSU). The phase modulator was driven by a 30 MHz sinusoidal voltage at a modulation depth of π. The output of the phase modulator was collimated by a lens and passed through two acetylene gas cells, each with a pass length of 200 mm. The first acetylene cell was filled with $^{13}C_2H_2$ (10 torr) and the second was filled with $^{12}C_2H_2$ (8 torr). The transmitted laser beams were divided by a band pass filter (BPF) that only transmitted light within a wavelength range of 1540–1550 nm and reflects light of other wavelengths. The divided beams are detected by amplified InGaAs photo detectors (PDA10CF, THORLABS). A dual-channel demodulation unit (PDD110, TOPTICA) produced error signals proportional to the laser frequency deviations from the absorption lines.

The optical frequency of the DFB-LD1 was controlled by a feedback controller (FALC110, TOPTICA), which fed the error signal back into the injection current of the DFB-LD1. Fig. 2 shows the measured frequency modulation response to the injection current of the DFB-LD. The phase delay was about 40 degrees at a modulation frequency of 200 kHz. This phase delay limited the bandwidth of the frequency control. Fig. 3 shows the frequency noise spectra of DFB-LD1 unlocked and locked to the acetylene absorption line. These spectra were measured by a feedback error signal with a spectrum analyzer (RSA3303B, Tektronix). The maximum unity-gain loop bandwidth was about 200 kHz, as expected from the phase delay of Fig. 2. Fig. 4 shows the frequency stability measured by the beat signal of DFB-LD1 and reference laser (C2H2LDS-1540, Neoark). The measured frequency stability with 1.0 sec averaging was 26 kHz with a standard deviation of 10 minutes.

A digital controller was used to control DFB-LD2 wavelength. The digital controller had two different wavelength control modes with closed loop control to the absorption line of $\lambda_2$ or $\lambda_3$ and open loop control for varying the wavelength between $\lambda_2$ and $\lambda_3$. A digital controller was configured by a PXI system (PXI7854R, National Instruments). The higher absolute distance measurement rate is enabled for faster wavelength scanning, but the wavelength scanning time was limited in practice by the frequency response characteristics of the DFB-LD shown in Fig. 2. The step response simulation with the measured frequency response suggested that more than 10 ms was required for converging to maximum capture range for closed loop control of 1 MHz.

Fig. 5 shows the wavelength scanning control results. The digital controller automatically alternated between closed loop control and open loop control at a rate of 1 Hz. The wavelength scanning time, which was defined as the time needed for converging the frequency error to 107 kHz, was 18 ms with a 5% overshoot for open loop control. These results show that the wavelength alternating rate can be increased to 50 Hz.

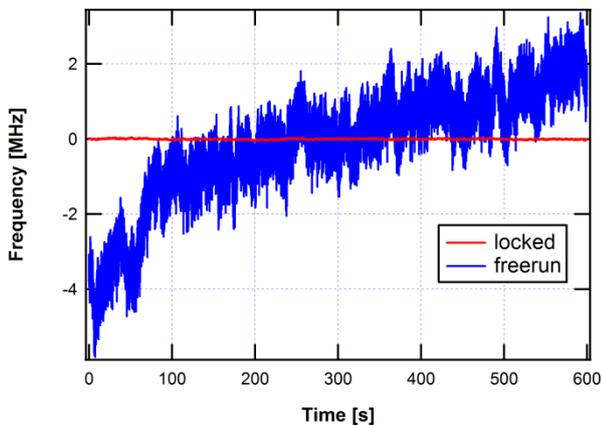

Fig. 4. (Color Online) Optical frequency stability of DFB-LD for freerun and locked conditions to the $C_2H_2$ absorption line within 10 minutes. The averaging time of each data was 1 sec.

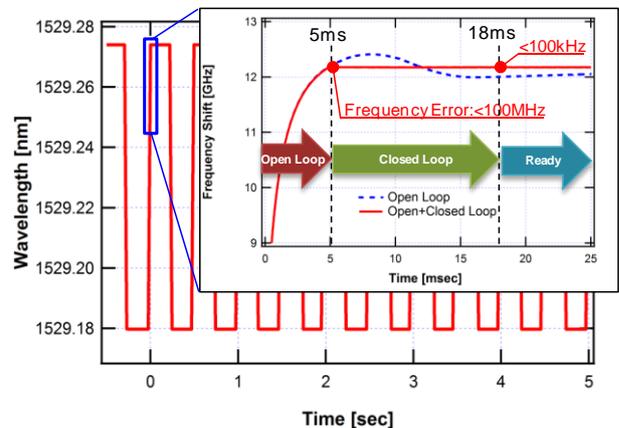

Fig. 5. (Color Online) Wavelength scanning control results. The wavelength scanning time, defined as the time for achieving a wavelength deviation less than 100 kHz, was 18 ms.

*Phase Detection Unit*

We adopted a heterodyne method for phase detection. The beam from a light source unit was delivered by a PM fiber and coupled to free space using a collimator. A polarization beam splitter (PBS) divided the beam into two orthogonal linear polarized beams. The optical frequency of one of the divided beam was shifted by 20 MHz using an acoustic optical modulator (AOM) (AFM-20-1540, BRIMROSE) and coupled to the other beam with a PBS. A non-polarizing beam splitter (NPBS) divided the beam into two parts, where 20% of the beam was used as a reference signal, and 80% was used as a measurement signal after passing through the interferometer unit. Both reference signal and measurement signal were delivered by a multi-mode fiber and divided into separate interference beams for DFB1 and DFB2 at BPF. Each beam was detected independently using amplified InGaAs detectors (PDA10CF, THORLABS). The detected signal was then digitized in the analyzer.

The analyzer calculated the unwrapped phase at 25 MHz by a custom algorithm on an FPGA. The unwrapped phase was output to a PC with a 25 kHz bandwidth and a 10 kHz data rate. Fig. 6 shows the phase accuracy test results, where the phase measurement data was obtained with optically equivalent reference and the measurement signals. The standard deviation was 0.6 m$\lambda$ with a phase resolution of 0.9 m$\lambda$.

*Interferometer Unit*

The interferometer was configured using two corner cubes for the reference and object mirrors. The object mirror was set in a linear stage an adjustable range of 1 m, enabling a change in optical path difference between the reference beam and the test beam from 0.5 m to 1.5 m. The displacement measurement interferometer (DMI) (ZMI4004/7702: ZYGO) was used for compensating environmental-induced displacement errors such as the variation of refractive index or the vibration of the object surface. Test beams of DMI and ADMi were coupled and collinearly aligned by a cold mirror for minimizing spatial air turbulence caused by refractive index error. The temperature dependence of the refractive index of air $dn/dT$ is 1 ppm/K. However, the difference in $dn/dT$ is only 0.01 ppm/K for the two measurement wavelengths of 633 nm and 1.5 μm. Therefore, less than $10^{-9}$ accuracy in distance evaluation was achieved under atmospheric environmental conditions with a temperature stability of 0.1 K.

One of the largest errors in interferometric measurements is known as cyclic error, induced by multiple interference of stray lighting. Fig. 7 shows the power spectrum density (PSD) of the displacement measurement error while the object surface was scanning linearly. The peak in PSD at 1.3 μm$^{-1}$, the reciprocal of the measurement period of 1.540/2 μm, is the cyclic error-induced noise spectrum. The cyclic error was estimated to be a standard deviation of 1.1 nm.

Some techniques exist for compensating the cyclic error of a heterodyne interferometer in real time [13,14]. Since we did not require real time characteristics in this demonstration, cyclic error was compensated by averaging position data with a small amount of stage stepping. We could reduce cyclic error by a factor of 1/20 by averaging 62 steps within a range of 3.1 μm, which was the least common multiple of the measurement wavelength of ADMi and DMI.

## 4. Measurement Result

First, the precision of a distance measurement using the ADMi is presented. The difference values of ADMi and DMI results were used for evaluation, since each interferometer shared parts of their optical paths, thus minimizing errors due to optical path variations. Since the measurement period of 10 ms was much shorter than a typical air turbulence cycle, optical path variations due to air turbulence were also minimized. Fig. 8 shows the evaluated precisions at three different distances from 0.5 m to 1 m. A standard deviation of 0.6 nm was achieved at a 0.5 m distance and increased at longer distances. Distance precision could be modeled as the sum of a distance independent component and a distance dependent component as:

$$\sigma_{Total}(L) = \sqrt{(\sigma_{phase})^2 + \left(\frac{\sigma_f}{f}L\right)^2}, \qquad (6)$$

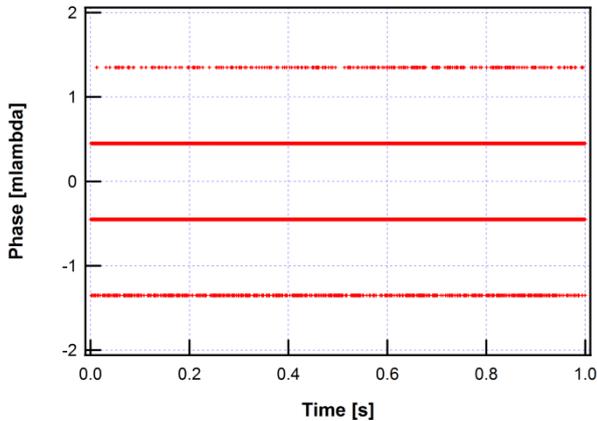

Fig. 6. (Color Online) Phase accuracy test result which was the phase measurement data in condition with the reference signal and the measurement signal are optically equivalent.

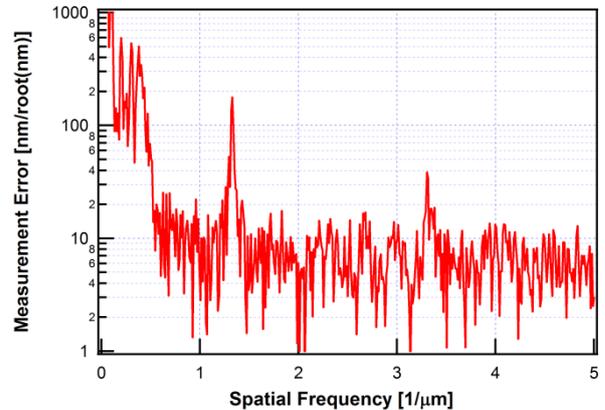

Fig. 7. (Color Online) Distance measurement error spectrum with linear stage scanning. The spectrum peak at 1.3μm$^{-1}$ shows a cyclic error of 1.1 nm in standard deviation.

where $\sigma_{phase}$ is the distance independent standard deviation due to phase measurement error and $(\sigma_f/f)L$ is the distance dependent component of the standard deviation due to the frequency error in the light source. We estimated the precision of these two components as 0.6 m$\lambda$ for $\sigma_{phase}$ and 132 kHz for $\sigma_f$ by least squares fitting of Eq. (6) to the experimental result of Fig. 6.

Second, we evaluated the accuracy of the absolute distance value from the ADMi. Before evaluation, we calibrated the three measurement wavelengths of the ADMi with the DMI wavelength as a standard. The interferometer unit was used as a scanning wavemeter during wavelength calibration. Measurement wavelengths used for the ADMi were calibrated such that the measured displacement values of the ADMi for the three wavelengths agreed with that from the DMI with stage scanning.

After wavelength calibration, we evaluated the accuracy of the interferometric orders $M$ and $N$ in Eq. (4). In order to determine $M$ and $N$ correctly, the arguments of the round function in Eq. (4) must be less than 0.5. The error in $M$ and $N$ are:

$$dM = \frac{n_{g12}}{n_{g23}}\frac{\Lambda_{23}}{\Lambda_{12}}(\phi_3-\phi_2)-(\phi_2-\phi_1)-n_{g12}\frac{L_{DMI}}{\Lambda_{12}}$$
$$dN = \frac{n_1}{n_{g12}}\frac{\Lambda_{12}}{\lambda_1}(M+\phi_2-\phi_1)-\phi_1-n_1\frac{L_{DMI}}{\lambda_1} \quad . \quad (7)$$

Fig. 9 shows the measured $dM$ and $dN$. Measurement data was time and position averaged in order to reduce cyclic error. The number of steps was 62 and the total measurement time for each data point was 1 sec. The maximum deviation of $dM$ and $dN$ was 0.13 for a distance range from 0.5 m to 1.5 m. Thus, the interferometric order could be determined properly.

Finally, we evaluate the accuracy of absolute distance values calculated using Eq. (3) and Eq. (4). Fig. 10 shows the difference in absolute distance values of the ADMi and relative displacement values of the DMI. The measurement condition was the same as the interferometric order evaluation. The accuracy of the absolute distance value in the distance range from 0.5 m to 1.5 m was estimated to be 1.2 nm in standard deviation, equivalent to a relative accuracy of $0.8 \times 10^{-9}$.

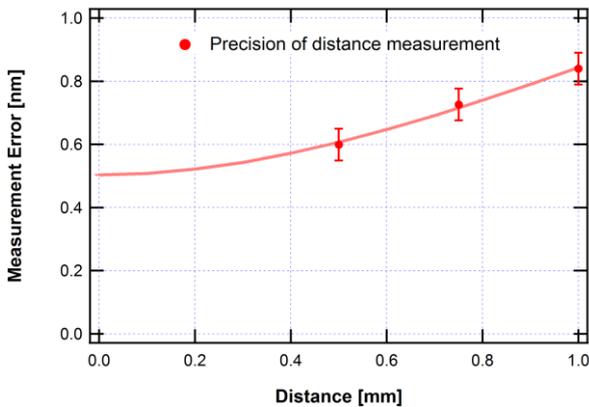

Fig. 8. (Color Online) Measured distance precision within 10 ms at distances from 0.5 m to 1.0 m.

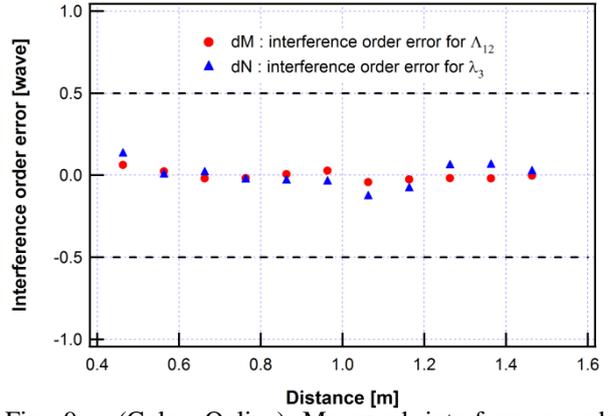

Fig. 9. (Color Online) Measured interference order error in the 1.5 m range. Each data point was taken with 64 step position averaging within the 3.1 μm range. Total acquisition time was 1 sec per position.

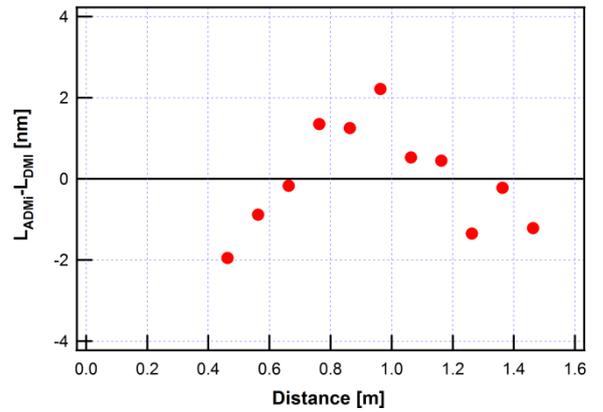

Fig. 10. (Color Online) The difference between absolute distance measurement results of the ADMi and relative displacement result of the DMI in the 1.5 m range.

## 5. For further improvement

We also show the two proposals for linear colliders. The first is a proposal to expand the range of maximum measurement distance up to 10 m, and the second is a proposal to compensate refractive index of air even in moist air.

### Expanding the range of maximum measurement

The error factors in Eq. (5) are the wavelength accuracies and the phase measurement accuracy. Because the wavelength accuracy term in Eq. (5) is proportional to $L$, the possibility to make a mistake in the interferometric order increases as measurement range becomes long. In this system, the wavelength was locked to the linear absorption lines of two acetylene isotopes. Generally a linewidth of linear absorption line is wide because of doppler broadening. The full width at half maximum of the linear absorption in this system is about 500 MHz [15], and it restricted the minimum wavelength stability to 100 kHz. Therefore, the measurement range of our system was limited to few meters.

However, it is possible to achieve higher wavelength accuracy by using an optical frequency comb (OFC) or a saturated absorption line as the frequency standard. The frequency stability of the OFC achieves 2 Hz [16], and that of the saturated absorption line is 10 kHz [17].

Therefore, the measurement range can be extended up to 10 m.

*Compensation of refractive index of air*
Laser interferometer can measure an optical path length, which is represented as the product of the refractive index of air and the geometric length. Therefore, in order to measure the geometric length with a high accuracy, we also have to measure refractive index with a high accuracy, and there have been a number of methods for compensation of refractive index.

There is a method to use environment parameter measurement and the empirical formula for the refractive index. As the empirical formula, Edlen's equation and Bönsch-Potulski's equation are well known [18,19]. According to Bönsch-Potulski's equation, refractive index of air $n$ is represented as:

$$n(\lambda, t, p, x, p_w) = 1 + K(\lambda) \times D(t, p, x) - p_w \times g(\lambda), \quad (8)$$

where t is the temperature, $p$ is the atmospheric pressure, $x$ is the CO2 content, $p_w$ is the partial pressure of water vapor, $D$ is the function which is dependent on $t$, $p$, and $x$, $K$ is the wavelength dispersion of the function $D$, and $g$ is wavelength dispersion of $p_w$. Therefore, it is possible to calculate the refractive index by measuring environment parameters by thermometer, hygrometer, barometer and others which are located near the optical axis, . This method has two problems. The first is that empirical formula itself has large uncertainty, and the second is that the positions of the environment parameter measurement cannot be located on the optical axis.

There is a method to measure refractive index directly instead of using the empirical formula [20]. In this method, a part of the light beam for the distance measurement is extracted, and enters an optical wavelength compensator arranged near the light beam for measurement of the distance. Here, the optical wavelength compensator consists of a reference pass with the evacuated tube and a test pass in air whose lengths are equal and already-known. We can calculate the refractive index by measuring the optical path length difference between the reference pass and the test pass, because these lengths are already-known. However, the refractive index measurement position cannot be located on the optical axis too, and the refractive index measurement accuracy depends on the spatial distribution of the refractive index.

A two-color method is known for decreasing the influence of the spatial distribution of the refractive index, and it achieves the accuracy of $5 \times 10^{-8}$ [21]. The principle of this method is to measure refractive index by using two different wavelengths. In this method, the geometric distance $L$ is:

$$L = OPL(\lambda_1) - A(OPL(\lambda_2) - OPL(\lambda_1)), \quad (9)$$

with

$$A = \frac{n(\lambda_1) - 1}{n(\lambda_2) - n(\lambda_1)}, \quad (10)$$

where, $OPL(\lambda_{i=1,2})$ is the optical path length by each wavelength. In the case of the dry air, the coefficient A is constant and independent on the environment parameter. Therefore, the accuracy of this method is very high even if the index measurement position cannot be located on the optical axis.

In the case of the moist air, the accuracy of the two-color method is reduced because the coefficient $A$ is not constant. Meiners et al achieved the accuracy of $1.2 \times 10^{-7}$ by measuring partial pressure of water vapor separately [22]. However, this method cannot compensate for the refractive index, if partial pressure of water vapor has spatial distribution. Therefore, we proposed the refractive index compensation system with high accuracy even if the moist air [23]. In this system, we added a wavelength of the absorption line of water vapor to the two-color method. Eq. (9) reduces to:

$$L = \frac{OPL(\lambda_2) - \alpha \times OPL(\lambda_1)}{1 - \alpha + p_w \times (\alpha \times g(\lambda_1) - g(\lambda_2))}, \quad (11)$$

with

$$\alpha = \frac{K(\lambda_2)}{K(\lambda_1)}. \quad (12)$$

Here, we can acquire partial pressure of water vapor pw in Eq. (11) by measuring the variation of laser intensity of the absorption line of water vapor. Thus, we can measure the average pressure of water vapor including the distribution of the pressure of water vapor on the optical axis. On the other hand, we can calculate α by the refractive index and partial pressure of water vapor in the optical wavelength compensator. Therefore, we'll achieve the refractive index compensation with high accuracy even in the moist air.

## 6. Conclusion

We demonstrated a diode laser-based absolute distance measurement interferometer with a 1.2 nm accuracy at a maximum measurement range of 1.5 m, yielding a relative accuracy of less than $10^{-9}$. This interferometer system had a simple design, composed only of two laser diodes, two gas cells filled with acetylene isotopes, and a two-wavelength heterodyne interferometer.

The wavelength scanning range required for absolute distance measurements by a varying synthetic wavelength interferometer could be decreased to 12 GHz using high accuracy phase detection of 0.6 mλ. High accuracy phase detection was accomplished with heterodyne detection for phase measurements for each wavelength of the interferometer and cyclic error compensation with stage scanning. The decreased wavelength scanning range enabled the use of the DFB-LDs as light sources for the interferometer and achieved a fast wavelength scanning time of 20 ms by injection current control. We achieved a wavelength stability of 100 kHz within 10 minutes by stabilizing to the linear absorption lines of the two acetylene isotopes.

Cyclic error is one of the main factors for interferometric phase measurement errors. There are some real time cyclic error calibration techniques based on the doppler shift of the measurement signal. Using these calibration techniques for our system, we can achieve the same accuracy without adding usage limitations such as position averaging or stage scanning.

We also proposed expanding the range of maximum measurement and compensation of refractive index of air for linear colliders. In this system, measurement range was mainly restricted by the wavelength accuracy of the linear absorption line. However, it is possible to achieve higher wavelength accuracy by using an optical frequency comb or a saturated absorption line as a frequency standard. Therefore, measurement range can be extended up to 10 m. In addition, the high accuracy of the ADMi is effective for the correction of refractive index. A relative accuracy of $10^{-7}$–$10^{-8}$ can be realized under atmospheric environmental conditions with the two-color method, known as the refractive index calibration method. However, this method cannot compensate for the refractive index, if partial pressure of water vapor has spatial distribution. Therefore, we proposed the refractive index compensation system in which we added a wavelength of the absorption line of water vapor to the two-color method. By measuring the variation of laser intensity of the absorption line of water vapor on the optical axis, we'll achieve the refractive index compensation with high accuracy even in the moist air.